\documentclass{article}
\usepackage{spconf,amsmath,graphicx}
\usepackage{color}
\usepackage{enumitem}
\usepackage{hyperref} 

\title{END-TO-END SOUND SOURCE SEPARATION CONDITIONED ON INSTRUMENT LABELS}
\name{Olga Slizovskaia$^{1 \dagger}$ \thanks{$^{\dagger}$ Equal contribution. Work done during the Deep Learning Camp Jeju 2018.}
 \qquad Leo Kim$^{2 \dagger}$ \qquad Gloria Haro$^{1}$ \qquad Emilia Gomez$^{1, 3}$}

\address{$^{1}$ Pompeu Fabra University \\
$^{2}$ University of Waterloo \\
$^{3}$ Joint Research Centre (EC)}
			    
\begin{document}
%
\maketitle
\begin{abstract}

Can we perform an end-to-end music source separation with a variable number of sources using a deep learning model? This paper presents an extension of the Wave-U-Net~\cite{waveunet} model which allows end-to-end monaural source separation with a non-fixed number of sources. Furthermore, we propose multiplicative conditioning with instrument labels at the bottleneck of the Wave-U-Net and show its effect on the separation results. This approach can be further extended to other types of conditioning such as audio-visual source separation and score-informed source separation.

\end{abstract}
\section{Introduction}
\label{sec:intro}

The goal of music source separation is to extract the mixture of audio sources into their individually separated source tracks. Undoubtedly, this is a challenging problem to solve and many attempts have been made to estimate the source signals as closely as possible from the observation of the mixture signals. The most common cases may vary with respect to the target task (such as singing voice \cite{faroit_overview, jansson2017singing} or multi-instrument source separation \cite{marius2016, pritish2017deepconv, oldorchestra}), use of additional information (blind \cite{pritish2017deepconv, jansson2017singing} or informed source separation \cite{marius2016, carabias2013nonnegative, oldorchestra}), and the amount of channels used for reconstruction (monaural \cite{pritish2017deepconv} or multi-channel \cite{marius2016, carabias2013nonnegative} source separation).   

There are many challenging aspects related to audio source separation. Most importantly, accurate separation with minimal distortion is desired. Supplementary information such as the number of sources present in the mix, musical notes in the form of MIDI or sheet music can be helpful but not widely available in most cases. However, information such as the source instrument labels can be easily found from video recordings of musical performances readily available on the web. Therefore, it sounds reasonable to learn to integrate the instrument label information into the source separation pipeline. At the same time, many sophisticated score- and timbre-informed methods have been proposed in the literature already \cite{faroit_overview}. We admire the idea of simplifying those frameworks, which became possible only recently with the advent of end-to-end deep neural networks.  

In this paper, we study how to separate musical recordings of small ensembles (from duets to quintets) into individual audio tracks. We propose an extension of the Wave-U-Net \cite{waveunet}, an end-to-end convolutional encoder-decoder model with skip connections, which supports a non-fixed number of sources and takes advantage of  instrument labels in assisting source separation.

\section{Related work}
\label{sec:background}

Traditionally, people have attempted to solve audio source separation through matrix-factorization algorithms. Independent Component Analysis (ICA) \cite{hyvarinen2000independent} and Non-negative Matrix Factorization (NMF) \cite{virtanen2007monaural} are two common techniques used for source separation. 


With the recent achievements in machine learning, researchers have started to adopt deep neural network paradigms to address the source separation problem. Since CNNs have been proven to be successful in image processing, raw audio data is often converted to 2D spectrogram images for analysis. The image data is then fed to a convolutional autoencoder which generates a set of masks that can be used to recover sound sources using inverse Short Time Fourier Transform (STFT) \cite{jansson2017singing, pritish2017deepconv, uhlich2017improving}.

In this paper, we aim to continue researching on deep learning methods for the source separation problem. Furthermore, we focus on improving the results by experimenting with less conventional approaches. Primarily, we work directly with raw waveforms as opposed to time-frequency image representation. This approach is an active research area \cite{waveunet, wavenetss2018} and gives us an additional advantage of preserving the phase information unlike other CNNs which only use the magnitude of STFT \cite{pritish2017deepconv, uhlich2017improving}. Secondly, we want to enhance our results through conditioning with instrument label information. This type of guidance has been shown to have a good impact on the source separation performance. Thus, in \cite{urmp_av2018}, the authors use visual guidance for improving source separation quality. Additionally, in a concurrent work~\cite{seetharaman2018class}, the authors explore a similar idea of class-conditioning over the joint embedded space, but unlike us, they use an auxiliary network to model parameters of a GMM for the final source separation, and they take spectrograms as an input of the model. 


Wave-U-Net model \cite{waveunet} is an adaptation of the U-Net \cite{unet}, a convolutional encoder-decoder network developed for image segmentation. The U-Net approach has been adapted already for singing voice separation in \cite{jansson2017singing}, however this model applies 2D convolutions and works with spectrograms. Instead of doing a 2D convolution, Wave-U-Net performs series of 1D convolutions, downsampling and upsampling with skip connections on a raw waveform signal. This approach was presented at SiSEC evaluation campaign \cite{sisec} and demonstrated competitive performance.

The input to this network is a single channel audio mix, and the desired output is the separated $K$ channels of individual audio sources, where $K$ is the number of sources present in the audio mix. 
 An interesting aspect of the Wave-U-Net is that it avoids implicit zero paddings in the downsampling layers, and it performs linear interpolation as opposed to de-convolution. This means that our dimension size is not preserved, and our output results will actually become a lot shorter compared to our input. However, by doing this we can better preserve temporal continuity and avoid audio artifacts in the results. 

\section{Expanded Wave-U-Net}
\label{sec:multisource}

\subsection{Multi-Source Extention}
The challenge with the original Wave-U-Net model is that it can only support a predefined number of input sources (2 and 4 sources in the original settings), limiting its application to only the specific group of instruments that it was trained on. We aim to build a more flexible model that can support a dynamic number of input sources and, therefore, be more suitable for separating classical music recordings.
In classical music, the number of instruments playing in an ensemble may vary a lot but the instruments themselves are often known in advance. Here we don't tackle the problem of separating different parts played by the same instrument (like violin1 vs violin2) but rather try to separate a sound track played by the same instrument (violin1+violin2 vs viola). 
Therefore, we can fix a maximum number of output sources to the number of all different instruments which are present in the dataset. 
This is still not a true dynamic model since the number of sources has to be specified in advance. 
Thus, in order to have a more general model we fix the number of sources to a reasonable large number.

For the sources that are not available in the mix, the model is trained with silent audio as a substitute. Therefore, the model outputs all possible sources and it is forced to associate each output with a certain instrument and output silence for the sources that are not present in the mix. Note that at the training time we implicitly specify which source should be aligned with a particular instrument, but it is not needed at the inference time. We can instead use an energy threshold for extracting the sources of interest. We will refer to this model as \textit{Exp-Wave-U-Net}.

\subsection{Label Conditioning}
 In order to enhance the source separation results, we propose a conditioned label-informed Wave-U-Net model (\textit{CExp-Wave-U-Net}). In particular, we use a binary vector whose size is the maximum number of sources considered. Each position of the vector is associated with a certain instrument: 1 indicates that the instrument is being played and 0 indicates either a non present instrument or a silent instrument (non-playing).

Conditioning is a term used to describe the process of fusing information of a different medium in the context of another medium. In case of Wave-U-Net, there are three locations where the use of conditioning is appropriate and corresponds to different fusion strategies: 
\begin{itemize}[topsep=1pt,itemsep=1pt,parsep=0pt,partopsep=1pt]
    \item for early fusion, the conditioning can be applied to the top layer of the encoder, before downsampling;
    \item for middle fusion, we can integrate label information at the bottleneck of the Wave-U-Net;
    \item for late fusion, we can aggregate labels with audio output of the last decoder layer (after upsampling).
\end{itemize}

Moreover, there is a possibility of using several conditioning mechanisms (as described in~\cite{dumoulin2018featurewise}) such as 
\begin{itemize}[topsep=1pt,itemsep=1pt,parsep=0pt,partopsep=1pt]
    \item concatenation-based conditioning;
    \item conditional biasing (additive bias);
    \item conditional scaling (multiplicative bias).
\end{itemize}

In this paper, we experiment with multiplicative conditioning using instrument labels at the bottleneck of the Wave-U-Net model. Therefore, the overall idea is to cancel out the unwanted sources at the most compressed part of Wave-U-Net while emphasizing the sources of interest. Even though the early fusion approach can be more abundant as it allows to integrate more information from the very beginning, we use multiplicative middle fusion as it provides a reasonable trade-off between expressiveness and memory and computational costs. At the same time, we leave additive bias and concatenation-based conditioning for further investigation.

\section{Implementation Details and Results}
\label{sec:implementation}
\subsection{Dataset}
As described in Sec.~\ref{sec:multisource}, the model takes the input in a form of a mix of the output sources where each source is either an instrumental track or a silent audio track for instruments not present in the mix. Instrument labels can be included optionally. We took advantage of the University of Rochester Musical Performance Dataset (URMP) \cite{urmpdataset} which consists of 44 pieces (11 duets, 12 trios, 14 quartets and 7 quintets) played by 13 different instruments (see Figure \ref{fig:res}). We used 33 pieces for training and validation, and 11 pieces for testing. 

\subsection{Baseline}
For the evaluation, we compare two proposed models with a Timbre-Informed NMF method from \cite{carabias2013nonnegative}. In this method, the authors first learn a timbre model for each note of each instrument, and apply this trained templates as the basis functions in NMF factorization procedure. Note that the timbre templates are trained with RWC \cite{Goto04RWC}, a dataset which consists of recordings of individual notes for different instruments.
Unlike our approach, Timbre-Informed NMF requires specifying the timbre models for each piece at the inference time. 
We used learned timbre models for all instruments except for saxophone.

\subsection{Implementation Details}

Our implementation is available online\footnote{\url{https://github.com/Veleslavia/vimss}} and is based on the original Wave-U-Net code\footnote{\url{https://github.com/f90/Wave-U-Net}}. We improved both input and training pipelines compared to the original work. The input pipeline is implemented as a TensorFlow Dataset and now supports parallel distributed reading. The training pipeline is re-implemented via a high-level TensorFlow Estimator API and supports both local and distributed training. Our implementation also supports half-precision floating-point format, which allows us to increase both training speed and batch size without loss of quality.

We train the model on a single Google Cloud TPU instance for 200k steps which takes approximately 23 hours. The best results are achieved using Adam optimizer with an initial learning rate of 1e-4. The aforementioned modifications together with the use of TPU allowed us to speed up training process by 24.8 times (35.3 times for the half-precision case) compared to a single GPU training. 

\subsection{Results}
\label{ssec:subhead}

We perform quantitative evaluation of the model performance using standard metrics for blind source separation: \textit{Source to Distortion Ratio} (SDR), \textit{Source to Inference Ratio} (SIR), and \textit{Source to Artifacts Ratio} (SAR) \cite{metrics}. 

\begin{table}[]
\begin{center}
\begin{tabular}{l|ccc}

\textbf{Method}  & \textbf{SDR} & \textbf{SIR} & \textbf{SAR} \\ \hline
 InformedNMF \cite{carabias2013nonnegative}  & \textbf{-0.16} & 1.42 & 9.31  \\ 
 Exp-Wave-U-Net & -4.12 & -3.06 & \textbf{12.18} \\ 
 CExp-Wave-U-Net & -1.37 &  \textbf{2.16} & 6.36 \\ 
\end{tabular}
\end{center}
\caption{URMP \cite{urmpdataset} dataset: SDR, SIR and SAR for different methods averaged over the testing set. Best values are shown in bold. Exp-Wave-U-Net states for an extension of Wave-U-Net with multiple output sources, CExp-Wave-U-Net states for a version of Exp-Wave-U-Net conditioned by labels of the instruments.}
\label{results}
\end{table}

\begin{table}[]
\begin{center}
\begin{tabular}{lc|ccc}
& &	\textbf{SDR} & 	\textbf{SIR}  & \textbf{SAR} \\
\textbf{Model}&	\textbf{nSources} & & & \\ \hline	
InformedNMF\cite{carabias2013nonnegative} &	2&	3.08&	4.98 & 10.55\\
&3&	0.07&	1.69&	9.01\\
&4&	-3.84&	-2.62&	8.65\\ \hline
Exp-Wave-U-Net&	2&	-0.42&	1.75 &	10.98\\
&3&	-3.85&	-2.74&	11.97\\
&4&	-5.90&	-5.33&	12.87\\ \hline
CExp-Wave-U-Net	&2&	-0.16&	4.62 &	7.48\\
&3&	-0.68&	2.88 & 5.91\\
&4&	-2.56&	0.44 &	6.35\\ 
\end{tabular}
\end{center}
\caption{SDR, SIR and SAR for different methods averaged with respect to the number of sources in the mix. }
\label{table:nInstrument_results}
\end{table}

\begin{figure}[!htb]

\begin{minipage}[b]{1.0\linewidth}
  \centering
  \centerline{\includegraphics[width=8.5cm,height=4.5cm]{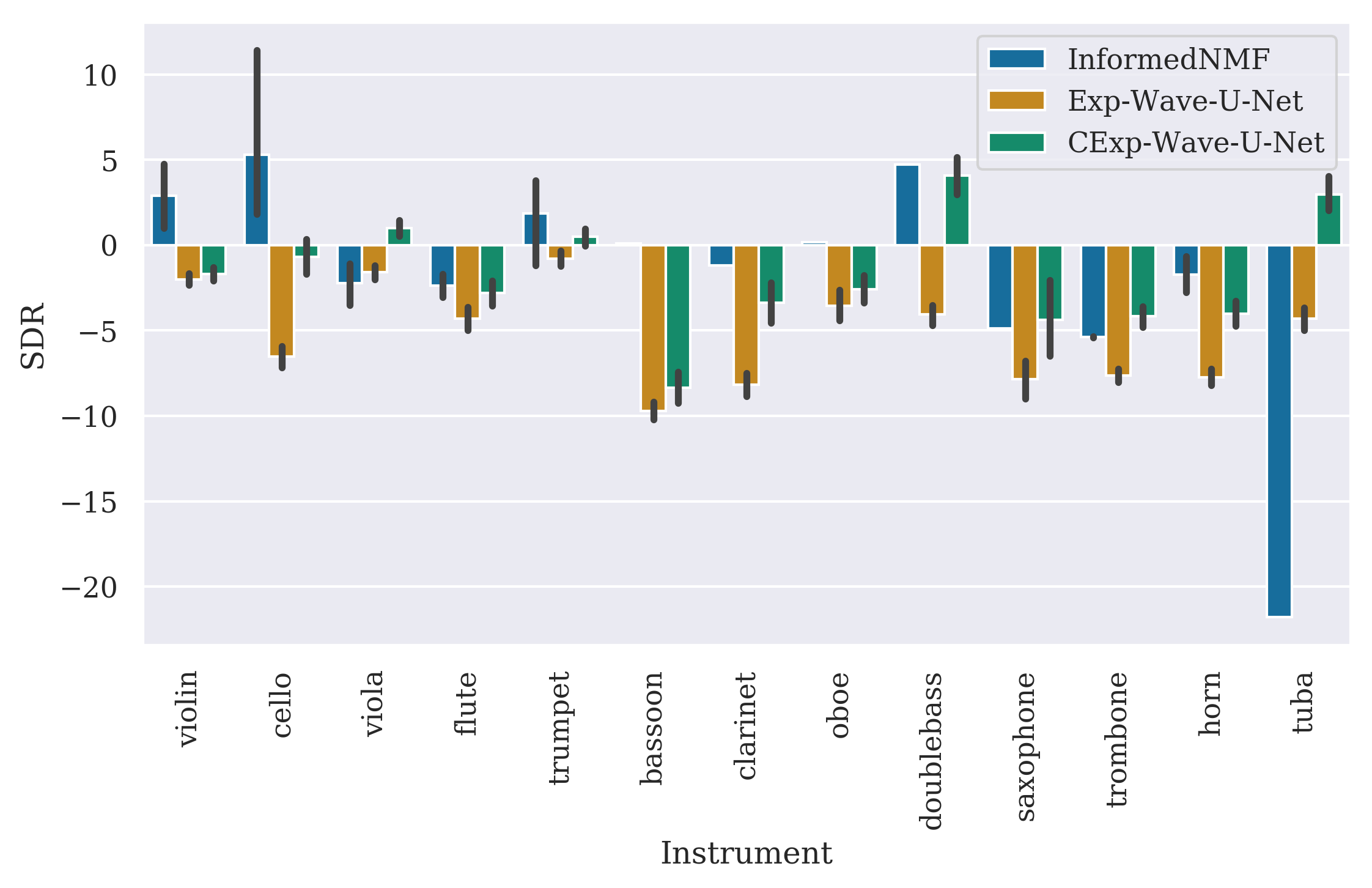}}
  \centerline{(a) SDR (dB)}\medskip
\end{minipage}
\begin{minipage}[b]{1.0\linewidth}
  \centering
  \centerline{\includegraphics[width=8.5cm,height=4.5cm]{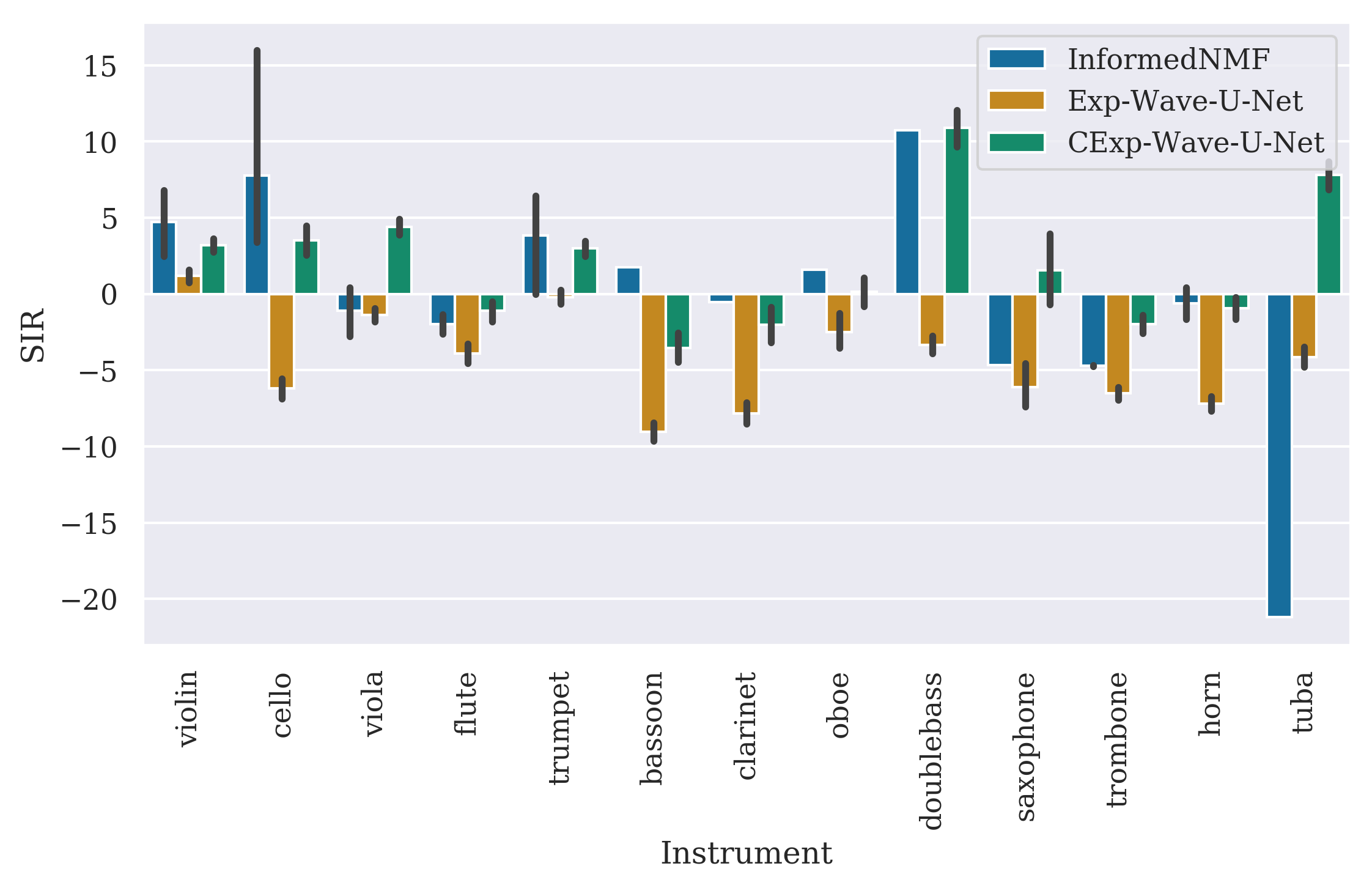}}
  \centerline{(b) SIR (dB)}\medskip
\end{minipage}
\hfill
\begin{minipage}[b]{1.0\linewidth}
  \centering
  \centerline{\includegraphics[width=8.5cm,height=4.5cm]{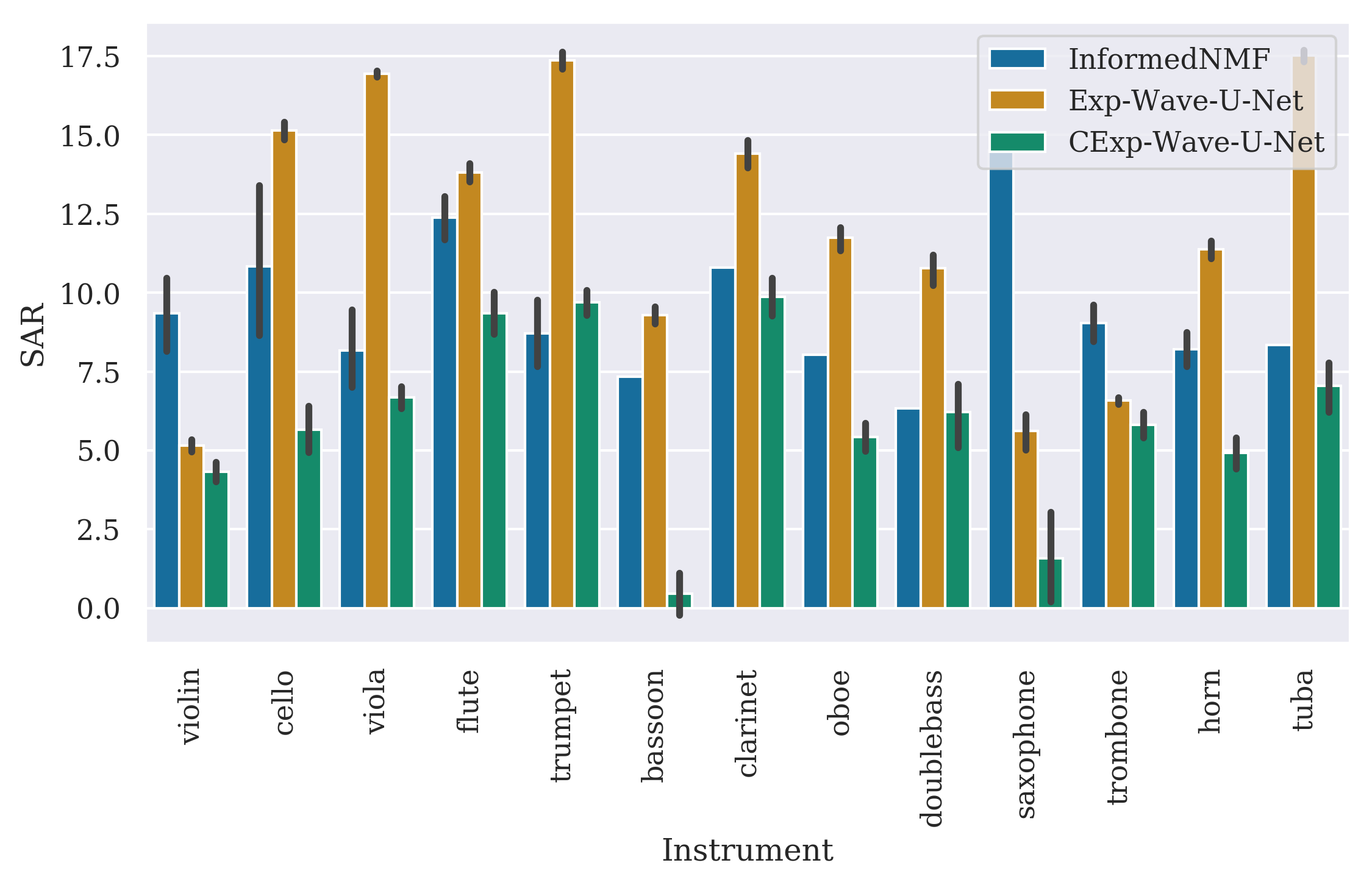}}
  \centerline{(c) SAR (dB)}\medskip
\end{minipage}
\caption{Results in terms of SDR, SIR, and SAR for each instrument in the testing set of URMP \cite{urmpdataset} dataset.}
\label{fig:res}
\end{figure}

\begin{figure}[!htb]

\begin{minipage}[b]{1.0\linewidth}
  \centering
  \centerline{\includegraphics[width=8.5cm]{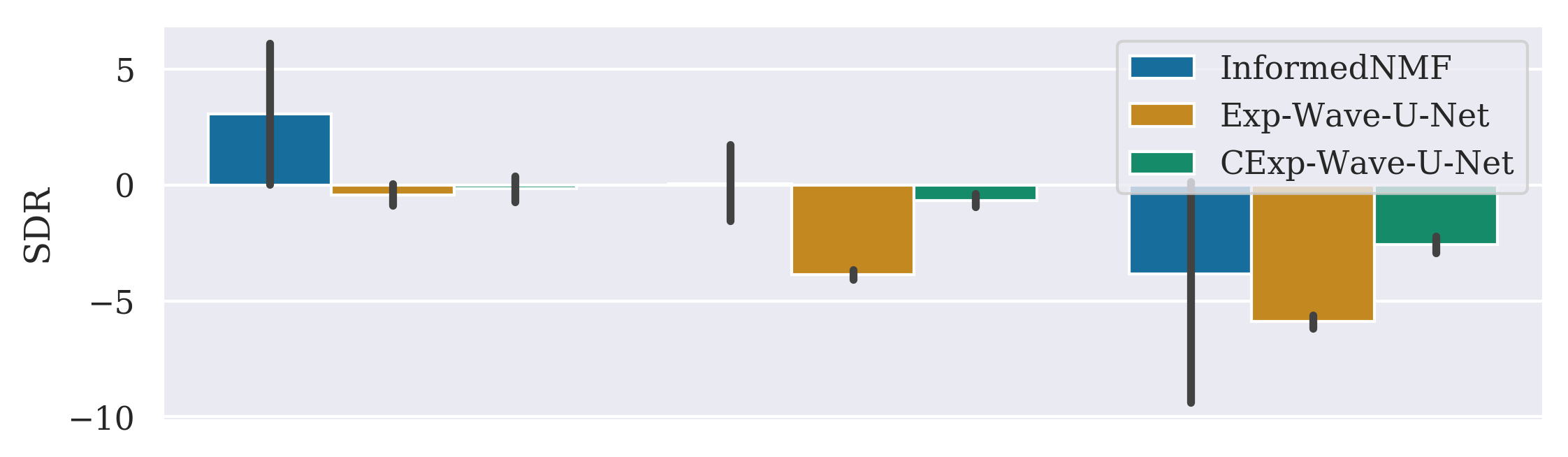}}
  \centerline{\includegraphics[width=8.5cm]{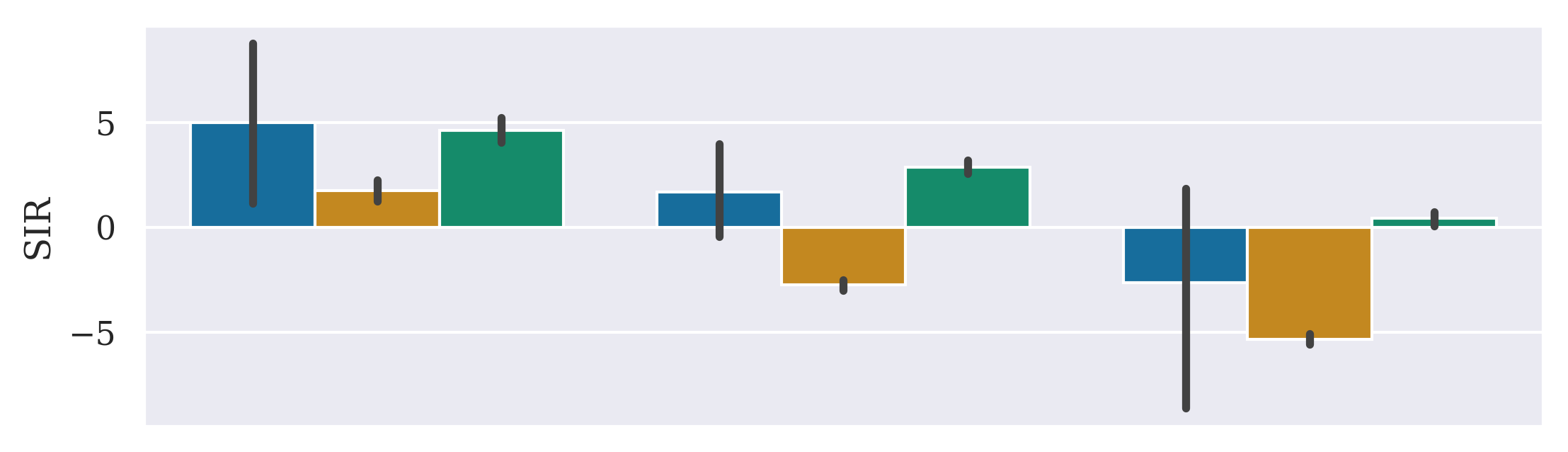}}
  \centerline{\includegraphics[width=8.5cm]{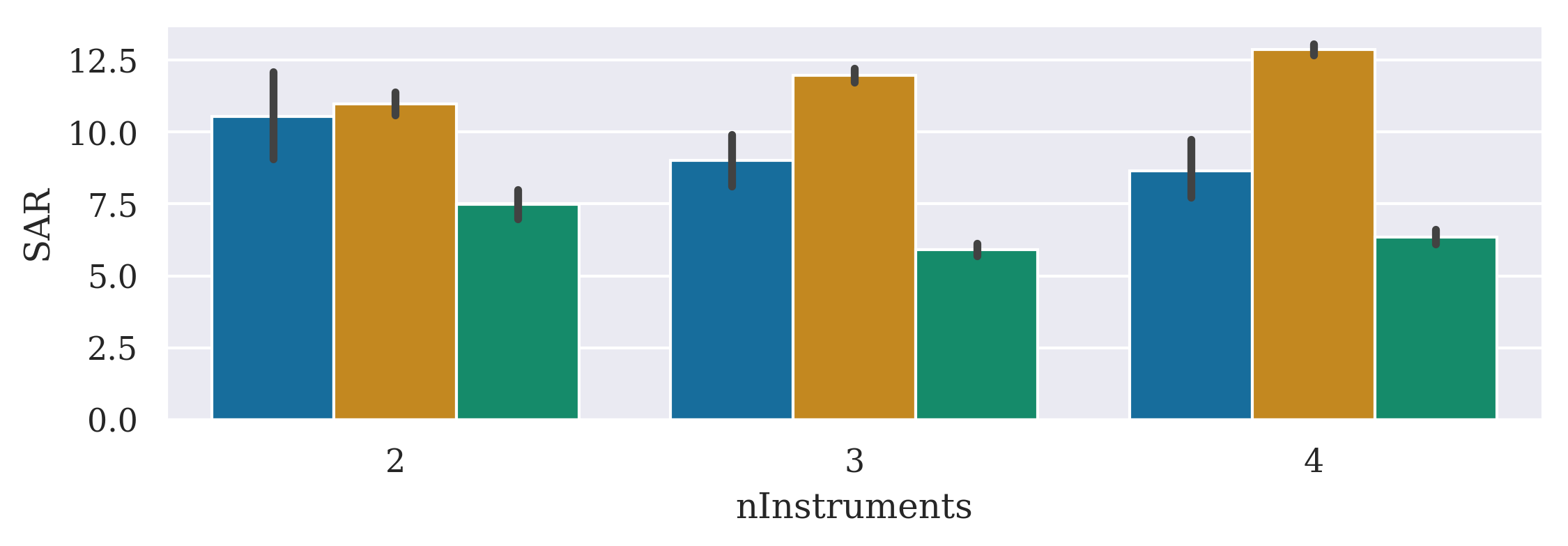}}

\end{minipage}

\caption{Results in terms of SDR, SIR, and SAR averaged and reported by the number of instruments in the testing set of URMP \cite{urmpdataset} dataset.}
\label{fig:res_ninstr}
\end{figure}

Table \ref{results} shows average values of the metrics over all pieces and instruments in the dataset. We can see that there is no single winner but each method seems to be better with respect to one of the metrics. For example, InformedNMF baseline outperforms both deep models in terms of SDR while it is inferior to Exp-Wave-U-Net in terms of SAR and to CExp-Wave-U-Net in terms of SIR. Note that we can't directly compare our results with Wave-U-Net because it would require to train from 3 to 11 different models while for Exp-Wave-U-Net we just train a single model for all instruments and number of sources.

Next, we analyze the separation performance in depth for each instrument. Figure \ref{fig:res} summarizes the results for each model and metric. We can see that the baseline approach (InformedNMF) performs reasonable well in terms of SDR and SIR for all instruments except for trombone and tuba. Exp-Wave-U-Net performs worse in SDR and SIR for all instruments but consistently outperforms the baseline and CExp-Wave-U-Net in SAR except for violin, trombone and saxophone. 
CExp-Wave-U-Net performs as good as the rest two in SDR and SIR (and achieves best results for tuba, doublebass, saxophobe and viola) but consistently worse in SAR. 

At last, we report the separation results averaged with respect to the number of sources in the input mix in Figure \ref{fig:res_ninstr}. It is interesting to note that the performance of all methods goes down as the number of sources increases. Hovewer, it is more interesting that the performance of CExp-Wave-U-Net does not drop as much as in case of InformedNMF and Exp-Wave-U-Net. In absolute values (see Table \ref{table:nInstrument_results} 
), SDR for CExp-Wave-U-Net decreases from -0.16 dB to -2.56 dB while for the model without conditioning those values are -0.42 dB to -5.90 dB, and from 3.08 dB to -3.84 dB for the NMF baseline. The alike behaviour persists for SIR. From these results, we could anticipate that the conditioned model is more suitable for multi-instrument source separation.

We would like to mention that despite their widespread use, the standard metrics are unable to estimate how well the model can discard unwanted sources (they are undefined if the ground truth is silence). Nonetheless, we would like to provide samples of separated sources which should be discarded\footnote{\url{https://goo.gl/e18F41}}. We notice that both conditioned and unconditioned versions of Exp-Wave-U-Net systematically output quieter sources for the absent instruments than InformedNFM, initialized by all possible timbre templates.

Some qualitative results for original\footnote{\url{https://youtu.be/mGfhgLt1Ds4}} and expanded\footnote{\url{https://youtu.be/mVqIMXoSDqE}} Wave-U-Net can be also found online.

\section{Conclusion}
\label{sec:conclusion}

In this paper we have proposed and explored two extensions of the Wave-U-Net architecture in the context of source separation of ensemble recordings with unknown number of input sources. We have shown that both Exp-Wave-U-Net and CExp-Wave-U-Net perform fairly competitive to the InformedNMF model despite being trained just on 33 audio mixes. We observed that CExp-Wave-U-Net outperforms the baseline approach when the number of input sources is bigger than 2. Moreover, we observed that Exp-Wave-U-Net produces a quieter output for the non-present instruments. 

We plan to further experiment with different fusion models for conditioning and incorporate visual information available within URMP dataset. The visual guidance seems to be a prominent direction of research because in this case not only we do not need to have manually annotated instrument labels but can also get an additional information of the playing and non-playing state of each instrument by analyzing the corresponding video stream. This can be especially useful to resolve disambiguation and inference between two instruments of the same kind.



{\bf Acknowledgements.} We would like to thank Terry Um and Eric Jang for their support during the camp, and Marius Miron for providing the code for the baseline. This work has received funding from the Maria de Maeztu Programme (MDM-2015-0502), ERC Innovation Programme (grant 770376, TROMPA), and MINECO/FEDER UE project (TIN2015-70410-C2-1-R).


\bibliographystyle{IEEEbib}
\bibliography{strings,refs}

\end{document}